\documentclass[pss]{wiley2sp}
\usepackage{amsmath}

\begin{document}

\title{Theoretical description of time-resolved pump/probe photoemission in TaS$_{\textsf{\bfseries 2}}$: a single-band DFT+DMFT(NRG) study within
the quasiequilibrium approximation}

\titlerunning{Theoretical description of time-resolved \ldots }

\author{%
  J. K. Freericks\textsuperscript{\textsf{\bfseries 1,\Ast}},
  H. R. Krishnamurthy\textsuperscript{\textsf{\bfseries 1,2,3}},
  Yizhi Ge\textsuperscript{\textsf{\bfseries 1}},
  A. Y. Liu\textsuperscript{\textsf{\bfseries 1}}, and
  Th. Pruschke\textsuperscript{\textsf{\bfseries 4}} }

\authorrunning{Freericks et al.}

\mail{e-mail
  \textsf{freericks at physics dot georgetown dot edu}, Phone
  +202-687-6159, Fax +202-687-2087}

\institute{%
  \textsuperscript{1}\,Department of Physics, Georgetown University, Washington, DC 20057 USA\\
  \textsuperscript{2}\,Centre for Condensed Matter Theory, Department of Physics, Indian Institute of Science,
 Bangalore 560012, India\\
  \textsuperscript{3}\,Condensed Matter Theory Unit,
Jawaharlal Nehru Centre for Advanced Scientific Research,
Bangalore 560064, India\\
  \textsuperscript{4}\,Institute for Theoretical Physics, University of G\"ottingen, Friedrich-Hund-Platz 1,
D-37077 G\"ottingen, Germany}

\received{XXXX, revised XXXX, accepted XXXX} 
\published{XXXX} 

\abstract{%
%
%
%
\abstcol{%
  In this work, we theoretically examine recent pump/probe photoemission experiments on the strongly correlated charge-density-wave insulator TaS$_2$.  We describe the general nonequilibrium many-body formulation of time-resolved photoemission in the sudden approximation, and then solve the problem using dynamical mean-field theory with the numerical renormalization group and a bare density of states calculated from density functional theory including the charge-density-wave distortion of the ion cores and spin-orbit coupling.
  }{%
  We find a number of interesting results: (i) the bare band structure actually has more dispersion in the perpendicular direction than in the two-dimensional planes; (ii) the DMFT approach can produce upper and lower Hubbard bands that resemble those in the experiment, but the upper bands will overlap in energy with other higher energy bands; (iii) the effect of the finite width of the probe pulse is minimal on the shape of the photoemission spectra; and (iv) the quasiequilibrium approximation does not fully describe the behavior in this system.}
}

%
%

\maketitle   

\section{Introduction}

Strongly correlated electronic systems are some of the most interesting and exotic systems in the quantum world.  They are compounds where the outermost electrons interact so strongly with each other that noninteracting band theory no longer applies, and they can exhibit novel quantum-mechanical effects.  Most experimental and theoretical techniques for strongly correlated electrons have been applied in equilibrium, or in the linear-response regime, where we have a wealth of knowledge and understanding about how these systems behave. Strongly correlated electron systems that are driven into a nonequilibrium state are much less understood, and represent a new frontier for the condensed-matter physics community.  One of the most important experimental techniques that allows access to the nonequilibrium regime are time-resolved pump/ probe studies, where an intense and ultrashort pulse of light hits the material, causing it to become excited into a nonequilibrium state, and then a much less intense but still ultrashort pulse is used to probe the system at different time delays after the pump pulse, to examine how the nonequilibrium dynamics lead the system to relax back to equilibrium.  In this work, we will focus on time-resolved pump/ probe photoemission experiments that have recently been completed in TaS$_2$~\cite{perfetti}.

\begin{figure}[tb]
\includegraphics*[height=\linewidth,angle=-90]{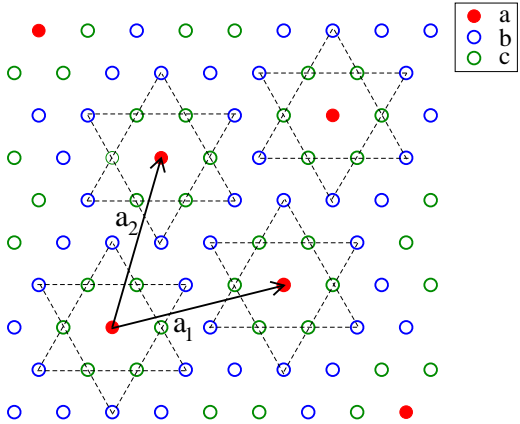}
\caption{(Color online.) Structure of the CDW-ordered Ta plane in the 1T$_3$ phase at low temperature.  There are three inequivalent Ta sites in the ordered phase, as indicated by the different colors.}
\label{struct}
\end{figure}

The material TaS$_2$ is an interesting strongly correlated commensurate charge-density-wave (CDW) insulator below an incommensurate/commensurate transition temperature of about 180~K~\cite{tas2_old}. Conventional band-structure calculations predict that the ordered phase is metallic, with a single band crossing the Fermi level, but experiment shows that the material is actually an insulator at low $T$, and it is believed the insulating state arises from Mott-Hubbard physics in this single band~\cite{fazekas} (which was originally proposed to be close to the critical $U$ for the Mott transition).  The CDW structure is actually quite complex in this system. Starting at temperatures above 1150~K, we find a (disordered) 1T structure is the stable structure for TaS$_2$; this structure is made up of S-Ta-S  trilayer units, in which atoms in each layer sit on a triangular lattice~\cite{Spijkerman}. 
Since the separation between
S-Ta-S sandwiches is larger than the interlayer spacing within a sandwich,
the structure is quasi two dimensional.
As the temperature is lowered, TaS$_2$ undergoes
a series of phase transitions to structures that can be viewed as distortions
of the basic 1T structure. Of interest here is the
low-temperature commensurate CDW phase (1T$_3$),
which appears below about 180 K.  In the plane, this phase is
characterized by a
$\sqrt{13}\times\sqrt{13}$ supercell in which the Ta atoms group into
clusters of 13 sites in  a  `Star-of-David' arrangement, as
shown in Fig.~\ref{struct}.
Some groups report that the  stacking sequence of S-Ta-S sandwich
units has a period of 13 \cite{Spijkerman}, while others suggest
that the stacking is disordered \cite{disordered}. For this work, we employ a $\sqrt{13}\times\sqrt{13}\times 1$
supercell to model the 1T$_3$ phase; i. e., we assume uniform stacking in the $z$-direction.

\section{Many-body formalism for TR-PES}

The theoretical description for time-resolved (TR) pump/probe photoemission spectroscopy (PES) is complicated.  For continuous beam photoemission, the signal is proportional to the so-called lesser Green's function (summed over $k_z$ if the system is not quasi two-dimensional) in frequency space, when one neglects the specific details of the matrix elements involved in the coupling of the electrons to the light and in the sudden approximation (see below for details).  For the nonequilibrium case, the formulation of the photoemission theory is more complicated.  One first must evolve the system in the presence of the pump pulse to create the `initial' nonequilibrium state on which the probe pulse acts. This is achieved by evolving the {\em equilibrium ensemble} of  many-body eigenstates $\{|\Psi_n\rangle\}$ of the system Hamiltonian $\mathcal{H}$ to the `initial' ensemble of states $\{|\Psi_n^I(t_0)\rangle\} \equiv \{U(t_0, -\infty)|\Psi_n\rangle\}$, where $U$ is the unitary time-evolution operator in the presence of the pump pulse, determined by a time dependent Hamiltonian $\mathcal{H}_{\rm pump}(t)$ whose precise form depends on the way one models the interaction of the pump radiation [represented by the the vector potential  ${\vec{A}}_{\rm pump}(\vec{r},t)$, with the $t$ dependence including the turning on and off of the pump,] with the electrons of the system~\cite{pump-model-ham}. The turning on of the probe pulse at time $t_0$ similarly modifies $\mathcal{H}$ by adding in the Hamiltonian $\mathcal{H}_{\rm probe}(t)$ which includes the interaction of the vector potential of the probe radiation with the system. At some later time $t$, (typically after the probe pulse has been turned off) the initial nonequilibrium ensemble of states $\{|\Psi_n^I(t_0)\rangle\}$ has evolved into the `final' ensemble of states $\{|{\Psi_n^F}(t)\rangle \} \equiv \{ {\tilde {U}}(t, t_0)|{\Psi_n^I}(t_0) \rangle \}$, with ${\tilde {U}}(t, t_0)$ being the unitary time-evolution operator that includes the presence of the probe pulse. The probability that a  photoelectron with momentum $\vec{k}_e \equiv k_e \hat{k}_e$ (in a momentum interval $dk_e$ and solid angle $d\Omega_{\hat{k}_e}$) is detected when the system is describable by this final nonequilibrium ensemble is  given by
\begin{equation}
\lim_{t\rightarrow\infty}\frac{(k_e)^2 dk_e d\Omega_{\hat{k}_e}} {(2 \pi)^3}P(t); \,P(t)\equiv \sum_{n,m} \rho_n \left |\langle{\Psi}_m ; \vec{k}_e| {\Psi_n^F}(t) \rangle\right |^2.
\label{PES-probability}
\end{equation}
Here, as appropriate to photoemission from an experimental sample {\em with a surface}, $|{\Psi}_m ; \vec{k}_e\rangle $ is well approximated as a direct product of the many-body eigenstate $|{\Psi}_m\rangle $ of the initial (and final) time-independent system Hamiltonian $\mathcal{H}$, and a {\it high energy, one-electron scattering state of the system} which propagates as a free electron state of momentum $\hbar\vec{k}_e$ outside the sample.  The eigenstate $|\Psi_m\rangle$ in which the system is left is not determined in the experiment, and the initial state can be any one of the ensemble of initial states with probability $\rho_n= \mathcal{Z}^{-1}\exp[-E_n/(k_B T)]$ where $E_n$ are the corresponding energy eigenvalues, and $\mathcal{Z}=\sum_n\exp[-E_n/(k_B T)]$ is the partition function. Hence Eq.~(\ref{PES-probability}) includes an unconstrained sum over $m$, and a sum over $n$ weighted by $\rho_n$. This expression implicitly assumes the sudden approximation, where the photo-excited electron rapidly moves out of the sample.

We employ a probe shape function, $s(t^\prime)$, to capture the time dependence of the temporal profile of the probe pulse, including its turning on and off. $M_{\vec{q}}(k_z, k_{ez};\vec{k}_\parallel)$, the matrix element for the absorption of a photon of wave vector $\vec{q}$~(frequency $\omega_{\vec{q}} = cq$) and the ejection of an electron of wave vector $\vec{k} \equiv (\vec{k}_\parallel, k_z)$ inside the system  as  a photo-electron of wave vector $\vec{k}_e = (\vec{k}_\parallel, k_{ez})$ depends on the details of the modeling of the sample, especially its surface~\cite{mat-el}.
To leading order in the perturbation $\mathcal{H}_{\rm probe}(t)$ due to the (weak) probe pulse (and using the factorization of $|{\Psi}_m ; \vec{k}_e\rangle $), we find
\begin {eqnarray}
\left |\langle {\Psi}_m ; \vec{k}_e| {\Psi_n^F}(t)\rangle \right | &\simeq&  \frac{1}{\hbar} \Big | \int dk_z M_{\vec{q}}(k_z, k_{ez};\vec{k}_\parallel) \nonumber\\
&\times&\int_{t_0} ^{t} dt^\prime s(t^\prime) e^{-i\omega t^\prime}\label{PES-amplitude}\\
&\times& \langle{\Psi}_m |U^{\dag}(t^\prime,t_0) c_{\vec{k}}U(t^\prime,t_0)|{\Psi_n^I}(t_0)\rangle\Big |,
\nonumber
\end{eqnarray}
with $U$ being the exact unitary time evolution operator in the presence of the pump pulse, which therefore includes all its nonequilibrium effects.
Here $ \hbar \omega \equiv  \hbar \omega_{\vec{q}} - $ \\
$(\hbar k_e)^2/(2m_e) - W $  is the energy of the excitation left in the system after the photoemission process and $W$ is the work function. For probe photons of fixed direction and energy, and for a given material, specifying $\omega$ determines $k_e$, hence the probability $P(t)$ in Eq.~(\ref{PES-probability}) is a function only of $t$, $\omega$ and $\hat{k}_e$. Using the properties of the time development operator, and the completeness of $\{|\Psi_m\rangle \}$  it is straightforward to show that
\begin{eqnarray}
P(t,\omega, \hat{k}_e) & \simeq & \frac{1}{(\hbar)^2} \int dk^\prime_z \int dk_z M_{\vec{q}}(k_z, k_{ez};\vec{k}_\parallel) \nonumber\\
&\times& M_{\vec{q}}^*(k^\prime_z, k_{ez};\vec{k}_\parallel) I(t,\omega, \hat{k}_e; k_z,k^\prime_z)
\end{eqnarray}
with
\begin{eqnarray}
I(t,\omega, \hat{k}_e; k_z,k^\prime_z) & \equiv & -i \int_{t_0} ^{t} dt^{\prime\prime} \int_{t_0} ^{t} dt^\prime s(t^{\prime\prime}) s(t^\prime) e^{i\omega(t^{\prime\prime}- t^\prime)} \nonumber\\
&\times&G_{\vec{k}, \vec{k}^\prime}^< (t^\prime,t^{\prime\prime}).
\label{eq: idef}
\end{eqnarray}
Here $\vec{k}\,' \equiv ({\vec{k}}_{\parallel},k^\prime_z)$ , and $G^<$ is the well known two-time (nonequilibrium) lesser Green's function~\cite{baym-kadanoff} given by
\begin{eqnarray}
G_{\vec{k},\vec{k}^\prime}^< (t^\prime,t^{\prime\prime}) & = & i \sum_n \rho_n \langle {\Psi}_n |U(-\infty, t^{\prime\prime}) c_{\vec{k}\,'}^{\dag} U(t^{\prime\prime},t^\prime) \nonumber\\
&\times& c_{\vec{k}} U(t^\prime, -\infty)| \Psi_n\rangle \\
&\equiv& i \mathcal{Z}^{-1}{\rm Tr}[e^{-\mathcal{H}/(k_B T)} c_{\vec{k}\,'}^{\dag} (t^{\prime\prime})  c_{\vec{k}}(t^\prime)]
\end{eqnarray}
where $c_{\vec{k}\,'}^{\dag} (t^{\prime\prime})$ and $c_{\vec{k}}(t^\prime)$ are the electron creation and destruction operators in the Heisenberg picture appropriate to $\mathcal{H}_{\rm pump}(t)$:
\begin{eqnarray}
c_{\vec{k}\,'}^{\dag} (t^{\prime\prime}) &\equiv& U(-\infty, t^{\prime\prime}) c_{\vec{k}\,'}^{\dag} U(t^{\prime\prime}, -\infty); \\
c_{\vec{k}} (t^\prime) &\equiv& U(-\infty, t^\prime) c_{\vec{k}} U(t^\prime, -\infty).
\end{eqnarray}
In the nonequilibrium (pumped) case, this needs to be calculated using nonequilibrium Keldysh contour-ordered Green's function techniques~\cite{Keldysh} on the Kadanoff-Baym-Keldysh contour in the complex time plane.

In case of continuous probe beam PES on a system in equilibrium,  $s(t) = 1$ and $G^<_{\vec{k}}(t^\prime,t^{\prime\prime})$ is only a function of  $(t^\prime-t^{\prime\prime})$. Furthermore, in a highly anisotropic layered system, its $k_z$ dependence can be neglected. Then the angle-resolved PES transition rate (transition probability per unit time, which is what is relevant as the continuous beam probe pulse photoemits electrons at all times at a constant rate) is proportional to
\begin{equation}
\lim_{t \rightarrow \infty}\lim_{t_0\rightarrow -\infty} \frac {I(t,\omega, \hat{k}_e)}{(t-t_0)} =  -i G_{{\vec{k}}_{\parallel}}^<(\omega) = A_{{\vec{k}}_{\parallel}}(\omega)f(\omega)
\end{equation}
which is the  standard result [with $A_{{\vec{k}}_{\parallel}}(\omega)$ being the spectral function and
$f(\omega)=1/\{1+\exp(\omega/k_BT)\}$ the Fermi-Dirac distribution function].

We will also invoke the quasiequilibrium approximation in this work.  This approximation assumes that the electronic system rapidly thermalizes after the pump pulse is turned off, but, because the time scale for thermalization with the phonons is much longer, the electronic system remains hot, and cools slowly during the course of the TR experiment~\cite{perfetti}.  In this case, the PES is described by the {\em equilibrium} lesser Green's function, but at an effective electronic temperature that depends on the time delay for the turning on of the probe pulse, convoluted with the probe pulse shape functions $s(t)$ as in Eq. \ref{eq: idef}.
Finally, we take the matrix elements $M$ to be constants, and evaluate the Green's functions for a bulk system, rather than a system with a surface, in which case we have full translational invariance, and $k_z=k_z^\prime$.
This last approximation is perhaps the most serious one that we make, but it also is commonly done in theoretical treatments of PES.

\section{Numerical techniques}

Because we need to describe Mott-Hubbard physics we will use a dynamical mean-field theory (DMFT) approach. Since there is only one band at the Fermi level, this problem can be formulated as a simple single band model with a complicated noninteracting DOS. (This tacitly assumes that any other higher bands remain higher in energy than the upper Hubbard band that comes from the original noninteracting band at the Fermi energy, a result that may well not be true in TaS$_2$ - see comments below.)  That DOS is determined via band structure calculations that include both the CDW distortions of the ion cores and the spin-orbit coupling.

The band structure calculations were done using the density-functional
code VASP \cite{vasp}, a plane-wave-based all-electron
code in which the electron-core interaction is treated using the
projector augmented wave method \cite{PAW}. The
generalized gradient approximation (GGA) was used for
electron exchange and  correlation \cite{PBE}.
For self-consistent-field calculations and structural relaxations,
the Brillouin zone was sampled using a $6\times6\times12$ grid
of {\bf k}-points.
Unless explicitly noted, calculations were carried out using the
scalar relativistic Hamiltonian.

The GGA optimized in-plane lattice
constant  of $a =  12.2$ \AA~ is about 0.5\% larger than the experimental
value.  For the out-of-plane lattice constant, the experimental
$c/a$ was used to mitigate effects arising from the simplified stacking
sequence  assumed.
Atomic positions were relaxed in the scalar relativistic approximation,
holding the lattice constants fixed.
The $b$- and $c$-site Ta atoms relax towards the center $a$ site,
with the $a$-$b$ and $a$-$c$
distances contracting by 5.3\% and 3.8\% from their respective values
in the  undistorted triangular lattice.
These results are close to the experimentally measured distortions.
For the S atoms, the relaxations were found to be larger in the $z$
direction than within the planes. The S planes develop a slight
pucker, with S atoms close to the central Ta $a$ sites displaced outwards
in the $z$ direction.

In the undistorted 1T structure,   a previous density functional
calculation has shown that the electronic bands have a
two-dimensional character in which the Ta $d$ bands crossing the Fermi
level are only weakly dispersing in  the $k_z$ direction, consistent
with the quasi 2D nature of the lattice \cite{Bovet}.
That study also found that the formation of the Star-of-David clusters
causes significant changes in the bands near the Fermi level. Indeed,
in our calculation, the CDW distortion causes the
Ta $d$ band that crosses the Fermi level to become essentially
one-dimensional, with a width of about 0.5 eV arising almost
exclusively from dispersion in the $k_z$ direction (see Fig. 2).
This is a nonbonding $d_{z^2}$ band
with primary  weight on the central Ta $a$ site.
The Ta clustering causes this state to become localized
in the in-plane directions, which suggests that the electronic structure
should transition from being two-dimensional to zero-dimensional.
However, the outward bulging of the S atoms around the central Ta $a$ sites
that accompanies the formation of clusters enhances the interactions between
tri-layer units, resulting in a one-dimensional band instead.

\begin{figure}[tb]
\includegraphics*[width=3.2in]{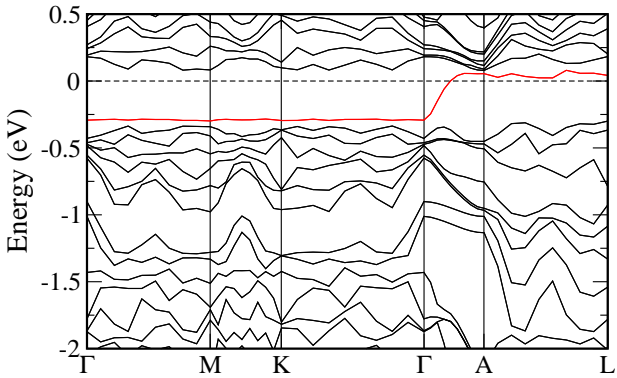}
\caption{(Color online.) Band structure of CDW-ordered TaS$_2$, including the effects of the
spin-orbit interaction. The zero of energy is set to the Fermi level. To
facilitate comparison with results in Ref. \cite{Bovet} which ignored
spin-orbit coupling, the bands are plotted along high-symmetry directions in the
Brillouin zone of the undistorted 1T structure.  The half-filled band
(red) that crosses the Fermi level  is strongly localized in the
in-plane directions, and is dispersive along $k_z$ ($\Gamma$ to A).}
\label{fig: bands}
\end{figure}

Recently, it has been suggested that in addition to
the CDW distortion, the spin-orbit effect could have  a large
influence on the bands near the Fermi level in TaS$_2$.
A  2D tight-binding model found that the spin-orbit
interaction causes the nonbonding Ta $d$ band
to split off from a manifold of higher-lying bands~\cite{Rossnagel}.
It becomes a very narrow (less than 0.1 eV in width)
half-filled band right at the Fermi level, primarily of
$d_{xy}$, $d_{x^2-y^2}$ character on the Ta $a$ site.
We have carried out density-functional calculations for the 3D system,
treating the spin-orbit effect self-consistently.  The results
are shown in Fig.~\ref{fig: bands}.
As in the tight-binding results, we find that the spin-orbit interaction
splits off the $a$-site nonbonding band from higher-energy bands.
However it still has a width of about 0.4 eV due to $k_z$ dispersion.
Its character remains mainly $d_{z^2}$, although
the  $d_{xy}$ and $d_{x^2-y^2}$  weight is significant at
the top of the band.

\begin{figure}[tb]
\includegraphics*[height=\linewidth]{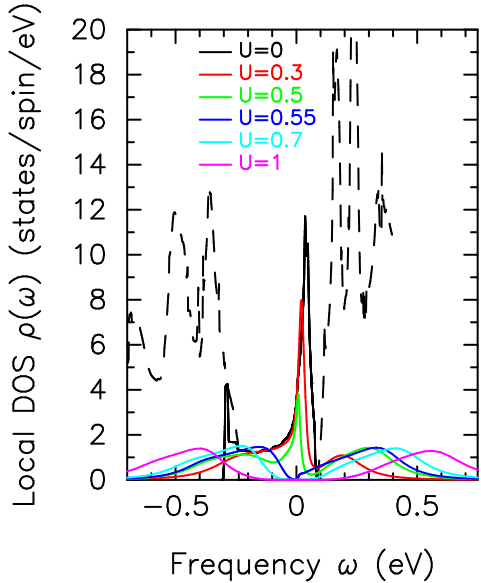}
\caption{(Color online.) Local many-body density of states for the single-band Hubbard model (at $T=50$~K) with the noninteracting
DOS given by the band structure calculations described in the text.  The metal-insulator transition takes place
around $U\approx0.55$~eV, and the best parameter value for TaS$_2$ is $U\approx 0.7$~eV. The dashed black lines show the upper and lower bands
in the band structure that lie above and below the band that crosses the Fermi energy; note how the upper and lower Hubbard bands merge with them
more and more as $U$ increases (even the noninteracting band is overlapping with neighboring bands near the lower band edge).}
\label{fig: dos}
\end{figure}

The density of states (DOS) of the split-off one-dimensional
band is shown in
Fig.~\ref{fig: dos} with the $U=0$ label. Its double-peaked shape can be understood from the $k_z$
dispersion: near the bottom of the band, the dispersion is quadratic,
leading to a DOS that goes roughly like $(E-E_0)^{-1/2}$;
near the middle of the band, the
dispersion is roughly linear, giving a nearly constant DOS;
and at the top of the band, the band is relatively flat in
all three directions, yielding a peak in the DOS.

In spite of the fact that the dispersion of the CDW-distorted bands (with spin-orbit coupling) have a quasi one-dimensional character to them, we nevertheless apply the DMFT approach to determine the Mott-Hubbard physics,
even if the results will only be approximate.
The DMFT calculations are now straightforward given the noninteracting DOS.  We make the simplifying assumption that the Coulomb interaction will be between the two different (degenerate) spin-orbit states at a given local site and use the DFT band structure DOS (for the single band that crosses the Fermi level) in the Hilbert transform that determines the local Green's function on the lattice from the momentum-dependent Green's function with a local self-energy. Then, the conventional numerical renormalization group (NRG) impurity solver can be used to solve for the Green's function of the effective impurity problem with the given value of the interaction $U$. The NRG approach we employ is the standard algorithm~\cite{nrg_review}---we take $\Lambda=1.8$ and keep 800 states per iteration on the Wilson chain. The DMFT equations for the retarded Green's function are solved via the self-consistent iterative approach of Jarrell~\cite{jarrell}. The PES signal is found by multiplying the spectral function by $-2if(w)$ to get the lesser Green's function, then Fourier transforming to real time to find $G^<(t-t^\prime)$.  Next we evaluate Eq.~(\ref{eq: idef}), which is proportional to the TR-PES spectra given the above assumptions.  We take the shape function $s(t)$ to be a Gaussian $s(t)=\exp[-(t-\bar t)^2/\Gamma^2]/(\Gamma\sqrt{\pi})$ with a width $\Gamma$ equal to 80~fs  and a probe delay time set at $\bar t$ (we assume the center of the pump pulse occurs at $t=0$). In the quasiequilibrium approximation, the lesser Green's function depends only on the time difference, so the convolution theorem can be used, and the final PES signal can be determined directly by a single integral in frequency space
\begin{equation}
 -i\int d\nu G_{{\vec{k}}_{\parallel}}^<(\omega-\nu)) \left |{\tilde s}(\nu)\right |^2/2\pi,
\end{equation}
where ${\tilde s}(\nu)$ is the Fourier transform of $s(t^\prime)$.  Since a pulse of 80~fs has a width on the order of 8~meV, this convolution will not modify the smooth structure in the DOS or the PES which vary over a frequency scale much larger than 0.01~eV.  Hence we neglect the ``windowing effect'' of the probe envelope function $s(t)$ in our analysis below. We fix our calculations to half-filling and vary the interaction strength to tune it for TaS$_2$, and then only vary the temperature.  Our PES signals are normalized so that the integrated weight in all spectra are identical.

We begin by calculating the local DOS at $T=50$~K to set the value of $U$. The results are shown in Fig.~\ref{fig: dos}.  Note how the metal-insulator transition takes place near $U=0.55$~eV. We ``fit'' to the experimental data by picking the distance between the peaks of the upper and lower Hubbard bands to be approximately 0.6~eV, as seen in the TR-PES studies~\cite{perfetti}. This result also produces a gap on the order of 0.125--0.15~eV, consistent with optical conductivity experiments~\cite{tas2_oc}. In addition, one should note that the upper and lower Hubbard bands are merging with the
next higher and next lower bands in the band structure, indicating that the single-band model may be inadequate for this system (particularly for the upper Hubbard band).

\section{Results}

The experimental continuous beam (equilibrium) PES studies on TaS$_2$ in the low-$T$ CDW phase show interesting behavior~\cite{grioni}.  The angle-resolved PES shows little dispersion of the lower Hubbard band peak as a function of momentum, consistent with the dispersion lying predominantly in the $k_z$ direction (the $k_z$ dispersion is seen in PES studies that vary the photon energy~\cite{Bovet}).  Furthermore, as one approaches the chemical potential, the angle-resolved PES curves tail off toward zero with an almost linear dependence on frequency.  This does not display the expected band gap in the insulating phase, where the spectra should go to zero before one hits the chemical potential.  It is argued in the experiment that this is occurring because the system has defects which pin the chemical potential to the upper edge of the lower Hubbard band.  If so, then there would be no way to see the gap in any continuous-beam (equilibrium) PES experiment, because one cannot reach the high temperatures needed before the CDW order disappears. In TR-PES experiments, one can access the higher bands by pumping energy into the system with the pump pulse which can excite electrons across the CDW gap and then be imaged.  But the TR-PES spectra also do not show a clear signature of a gap in the data, as we illustrate below.

\begin{figure}[tb]
\includegraphics*[height=\linewidth]{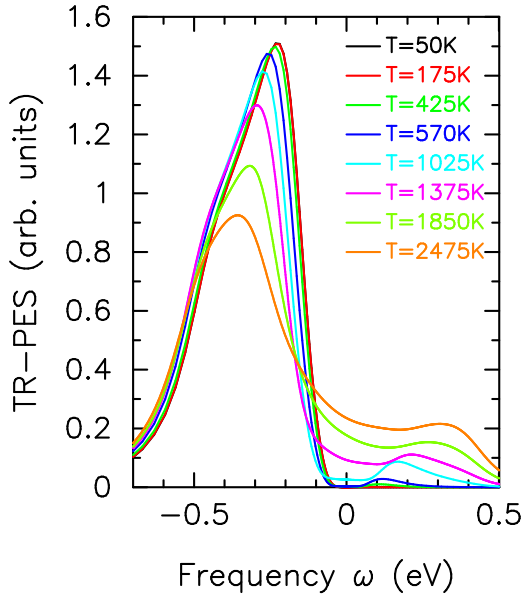}
\caption{(Color online.) Time-resolved photoemission spectroscopy in the single-band quasiequilibrium approximation for TaS$_2$.  Here we plot the local lesser Green's function to represent the sum over all $k_z$ values with the quasi one-dimensional dispersion.  Note how one always sees a strong insulator, and that one can image the upper Hubbard band as the temperature becomes on the order of half the bandwith and above. Note further that the 50~K curve and the 175~K curve are indistinguishable.}
\label{fig: pes}
\end{figure}

The TR-PES measurements on TaS$_2$ work with low-energy photons~\cite{perfetti}, and hence the electronic excitations are limited to lie in a small volume about the zone-center in the Brillouin zone. It is argued, if the system is quasi two-dimensional, that one would then examine the ${\bf k}=0$ spectral function.  We will show some results with just such an approach, although, the quasi one-dimensional dispersion along $k_z$, indicates that one should average instead over $k_z$, which is approximately the local Green's function when the dispersion is quasi-one-dimensional. So we also examine the local Green's function with our simplified approach.

\begin{figure}[tb]
\includegraphics*[height=\linewidth]{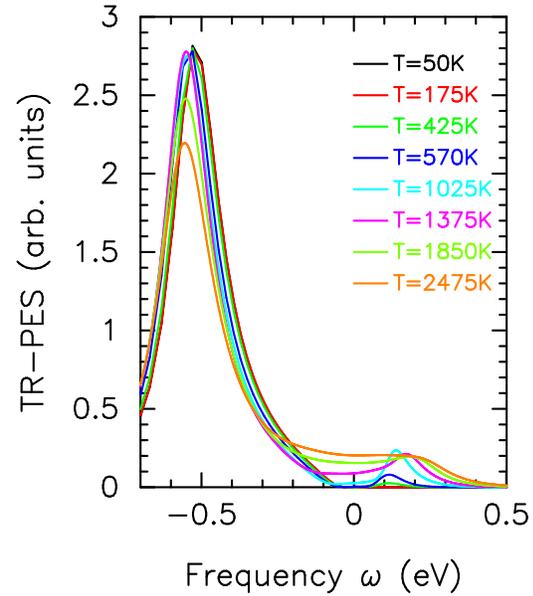}
\caption{(Color online.) Time-resolved photoemission spectroscopy in the single-band quasiequilibrium approximation for TaS$_2$.  Here we plot the $k=0$ lesser Green's function.  In this case, the relative weight of the upper and lower band features are closer to experiment than the local results and the tailing off of the signal as we approach the chemical potential from negative energies is also similar to what is seen in experiment, but the peak location is pushed much lower in energy and sits near $-0.5$~eV rather than $-0.2$~eV, as seen in experiment.}
\label{fig: pes2}
\end{figure}

Since we have set the Hubbard model interaction at $U=0.7$, we can determine the TR-PES spectra by simply calculating the local (or $k=0$) equilibrium lesser function at different temperatures and plotting as a function of frequency.  Different effective temperatures correspond to different time delays, with the hottest temperatures being the shortest time delays. The results of just such a calculation are shown for the local Green's function in Fig.~\ref{fig: pes} and for the $k=0$ Green's function in Fig.~\ref{fig: pes2}.  Here we plot the lesser function for different quasiequilibrium temperatures.  The chemical potential is fixed by the requirement that the filling remain exactly one in the band.  The chemical potential shifts in such a way to push the peaks to the left in the figures for higher temperature.  Such a shift is not seen in the experimental data.  If we instead assume that the chemical potential is pinned by defect states, then we would shift the curves back to the right (by hand) so they all meet at a characteristic frequency.  This is the behavior seen in experiment and could be reproduced in the theory if we allow ourselves to shift the frequency axis for each case or if we include defect states in our analysis.  In addition to the movement of the peaks with the changing chemical potential, we also clearly see insulating behavior for small $T$, and we always see a double peak structure when we can image the lower and upper Hubbard bands separately at high $T$. Note that the upper Hubbard band is always small in weight compared to the lower band due to the Fermi factor. The experimental TR-PES data does not show such behavior. It instead has an almost monotonically decreasing signal as one moves across the chemical potential and beyond, with no real gap structure to be seen anywhere.  This behavior is difficult to reconcile with a single-band Mott-Hubbard type of picture. The behavior for some of these features is reproduced better in the local approximation than the $k=0$ approximation (like the location of the peak height), while other behavior is produced better in the $k=0$ approximation (like the tailing off of the signal as one approaches the chemical potential, the lower relative weight of the upper Hubbard band, and the difficulty in directly imaging the peak in the upper Hubbard band).

We have a couple of conjectures about what is happening in this system, but we do not yet have definitive answers to these questions.  First off, the single-band model approximation is probably the most serious one that we made in this work.  This is because the higher-energy bands in the band structure lie just slightly above the noninteracting DOS upper band edge.  So the upper Hubbard band, which is pushed upwards in energy relative to the noninteracting band edge, is expected to lie in or above those first lower bands (the situation is not quite as bad for the $k=0$ spectra, because the $k=0$ band energies are pushed a bit higher).  This implies the system looks more like a charge-transfer insulator than a Mott-Hubbard insulator, and the upper bands in the band structure must be taken into account in the analysis.  If we instead ignore that problem, then we still need to explain why we do not see a clear signature of the upper Hubbard band in the data.  Here there are three other effects that can be playing a role.  One could have a situation where the nonequilibrium behavior is not described well by the quasiequilibrium approximation, and hence the ``filling in'' of the gap is a nonequilibrium effect.  Another possible explanation could arise from surface states which may lie at slightly different energies and could enhance the signal in the gap, making it look like there is no gap, when there actually is one in the bulk. Finally, the energy and momentum dependence of the matrix elements could be playing a role.  When one compares the TR-PES studies with the higher-energy continuous beam studies, one can see a difference in the relative peak heights of the lower Hubbard band and of the even lower energy filled bands, which must arise from matrix-element effects.

\section{Conclusions}
In this work, we have examined a theoretical description of the time-resolved photoemission spectroscopy in the strongly correlated CDW insulator 1T-TaS$_2$ based on a simplified single-band model that employs DFT + DMRG(NRG) techniques. The model can be solved directly, and does display some of the qualitative features of the experiments, but misses a number of key elements:  First, the experiments never show a true insulator, while the theory shows well defined insulating behavior even at high temperature, where a lower and upper Hubbard band can still be seen. Second, this approximation is neglecting the fact that the upper Hubbard band lies at energies that are typically above the lower band edge of the higher energy unoccupied band states of the system. When this occurs one really needs to consider a multiband model. Third, we find the DFT-based bands, with the CDW distortion and the spin-orbit coupling included, display quasi one-dimensional behavior perpendicular to the planes, and essentially localized behavior within the planes, indicating that averaging over $k_z$ is always necessary (even though many features in the experiment are represented better by the $k=0$ results with no $k_z$ averaging).  It is also possible that one need to take into account nonequilibrium effects beyond the quasiequilibrium approximation (which may fill the gap in the DOS faster), include the momentum or energy dependence to the matrix elements rather than treating them as constants, and incorporate surface effects, in particular the effects of surface states which could also mask the insulating behavior. Clearly further work is required to start to resolve and better understand 1T-TaS$_2$.

\section{Acknowledgments}

J. K. F., A. Y. L, and Y. G. acknowledge support from the National Science Foundation under grant number DMR-0705266.
H. R. K. is supported under ARO Award W911NF0710576 with funds from the DARPA OLE Program.
Th. P. acknowledges support from the collaborative research center (SFB) 602.
We also acknowledge useful discussions with
T. Devereaux,
M. Grioni,
B. Moritz,
L. Perfetti, and
K. Rossnagel.

\end{document}